\documentclass[12pt]{article}
\usepackage{qMobius}
\begin{document}
\title{Quantum Circuit for Calculating\\
Mobius-like Transforms\\
Via Grover-like Algorithm}

\author{Robert R. Tucci\\
        P.O. Box 226\\
        Bedford,  MA   01730\\
        tucci@ar-tiste.com}

\date{\today}

\maketitle
\vskip2cm
\section*{Abstract}
In this paper,
we give quantum circuits for
calculating two closely related linear transforms
that we refer to jointly as Mobius-like transforms. The first is the Mobius transform
of a function $f^{-}(S^-)\in \CC$,
where $S^-\subset \{0,1,\ldots,n-1\}$.
The second is a marginal
of a probability
distribution $P(y^n)$,
where $y^n\in Bool^n$.
Known classical algorithms for
calculating these Mobius-like transforms
take $\calo(2^n)$ steps.
Our quantum algorithm
is based on a Grover-like algorithm
and it takes
$\calo(\sqrt{2^n})$ steps.
\newpage

\section{Introduction}
In this paper,
we give quantum circuits for
calculating two closely related linear transforms
that we refer to jointly as Mobius-like transforms. The first is the Mobius transform
of a function $f^{-}(S^-)\in \CC$,
where $S^-\subset \{0,1,\ldots,n-1\}$.
Mobius transforms are defined
in Eq.(\ref{eq-mob-def}).
The second is a marginal
of a probability
distribution $P(y^n)$,
where $y^n\in Bool^n$.

Known classical algorithms for
calculating a Mobius transform
take $\calo(2^n)$ steps
(see Refs.\cite{KS,Ko-So}).
Our quantum algorithm
is based on the original Grover's algorithm (see Ref.\cite{Gro})
or some variant thereof (such as AFGA, described in Ref.\cite{afga}),
and it takes
$\calo(\sqrt{2^n})$ steps.

This paper assumes that
the reader has already read
most of Ref.\cite{qSym}
by Tucci. Reading that
previous paper is essential
to understanding this one
because
this paper applies
techniques described in
that previous paper.

\section{Notation and Preliminaries}

Most of the notation that will be
used in this paper has already been
explained in previous papers by Tucci.
See, in particular, Sec.2
(entitled ``Notation and Preliminaries") of Ref.\cite{qSym}.
In this section, we will discuss some
notation and definitions that
will be used in this paper but
which were not discussed in Ref.\cite{qSym}.

For any set $S$, let $2^S$
 represent its power set.
In this paper, we wish to consider a
finite set $S$, and two functions
$f, f^-:2^S\rarrow \CC$
related by

\beq
f(S) = \sum_{S^-\subset S}f^-(S^-)
\;.
\label{eq-mob-def}
\eeq
The sum in Eq.(\ref{eq-mob-def})
is over all subsets $S^-$ of
the ``mother" set $S$
(i.e., all $S^-\in 2^S$).
Function $f$ is called
the {\bf Mobius transform} of
function $f^-$.

Without loss of generality,
we may assume that
$S=\{0..n-1\}$.
If
$S^-\subset S$,
then we can write
$S^-=\{0^{x_0}, 1^{x_1}, 2^{x_2},\ldots,(n-1)^{x_{n-1}}\}$,
where $x^n\in Bool^n$.
In this notation for $S^-$,
if $x_j=0$, we are to omit
from the set $S^-$
the number being
exponentiated (the base),
whereas if $x_j=1$, we are to include it.
This notation for $S^-$ establishes a
bijection between $Bool^n$
and $2^{\{0..n-1\}}$.
Henceforth, we'll denote the two directions
of that bijection by
$S=S(x^n)$ and $x^n = x^n(S)$.
If $y^n,x^n\in Bool^n$,
define $y^n\leq x^n$ (or $x^n\geq y^n$) iff $(\forall j)(y_j\leq x_j)$. Clearly, $y^n\leq x^n$ iff $S(y^n)\subset S(x^n)$.

An equivalent way of
writing Eq.(\ref{eq-mob-def}) is

\beq
f(x^n) =
\sum_{x^{-n}\leq x^n}
f^-(x^{-n})
=
\sum_{x^{-n}\in Bool^n}
\theta(x^n\geq x^{-n})
f^-(x^{-n})
\;.
\label{eq-mob-def-xn}
\eeq
Note that

\beq
\theta(x^n\geq x^{-n})=
\prod_{j=0}^{n-1} \theta(x_j\geq x^-_j)
\;.
\eeq
For
$x, x^-\in Bool$,
define the matrix $M$ by

\beq
M =
\begin{array}{c|cc}
&\scriptstyle{x^-=0}& \scriptstyle{x^-=1}
\\ \hline
\scriptstyle{x=0} & 1 & 0
\\
\scriptstyle{x=1} & 1 & 1
\end{array}
\;,\;\;
M_{x,x'} =\theta(x\geq x^-)
\;.
\eeq
For
$x^2=(x_1,x_0)\in Bool^2$ and
$x^{-2}=(x^-_1,x^-_0)\in Bool^2$,
the 2-fold tensor product of $M$ is

\beq
M^{\otimes 2} =
\begin{array}{r|cccc}
&\scriptstyle{x^{-2}=00}& \scriptstyle{01}
&\scriptstyle{10}& \scriptstyle{11}
\\ \hline
\scriptstyle{x^{2}=00} &1&0&0&0
\\
\scriptstyle{01} &1&1&0&0
\\
\scriptstyle{10} &1&0&1&0
\\
\scriptstyle{11} &1&1&1&1
\\
\end{array}
\;,\;\;
(M^{\otimes 2})_{x^2,x^{-2}} =
\prod_{j=0,1}\theta(x_j\geq x^-_j)
\;.
\eeq
In general, for $x^n,x^{-n}\in Bool^n$,
the $n$-fold tensor product of $M$
is given by

\beq
(M^{\otimes n})_{x^n,x^{-n}} =
\prod_{j=0}^{n-1}\theta(x_j\geq x^-_j)
\;.
\label{eq-mn}
\eeq
From Eq.(\ref{eq-mn}),
we see that Eq.(\ref{eq-mob-def-xn})
can be written in matrix form as:

\beq
\ket{f} = M^{\otimes n} \ket{f^-}
\;,
\eeq
where $f(x^n)=\av{x^n|f}$
and $f^-(x^{-n})=\av{x^{-n}|f^-}$.

\section{Quantum Circuit For
Calculating Mobius Transforms}

In this section,
we will give a quantum
circuit for calculating the
Mobius transform of a probability
distribution $f^-(x^{-n})$
where $x^{-n}\in Bool^n$.
Our algorithm can also
be used to find the Mobius transform
of more general functions
using the method given in Appendix C
 of Ref.\cite{qSym}.

For $x^n,x^{-n}\in Bool^n$,
and a {\it normalized} $n$-qubit state $\ket{\psi^-}$,
define

\beq
\ket{\psi^-}_{\alpha^{-n}}=
\sum_{x^{-n}}
A^-(x^{-n})\ket{x^{-n}}_{\alpha^{-n}}
\;,
\eeq

\beq
f^-(x^{-n}) = |A^-(x^{-n})|^2
\;,
\eeq

\beq
f(x^n) =\sum_{x^{-n}\leq x^n}
f^-(x^{-n})
\;.
\eeq
Note that this function $f^-$
is not completely general. It's
non-negative and

\beq
f(1^n) = \sum_{x^{-n}}f^-(x^{-n})=1
\;.
\eeq

We will assume that
we know how to compile
$\ket{\psi^-}_{\alpha^{-n}}$
(i.e., that
we can construct it starting
from $\ket{0^n}_{\alpha^{-n}}$
using a sequence of
elementary operations.
Elementary operations are
operations that act on a few (usually 1,2 or 3)
qubits at a time,
such as qubit rotations
and CNOTS.)
Multiplexor techniques for doing
such compilations
are discussed in Ref.\cite{tuc-multiplexor}.
If $n$
is very large,
our algorithm will
be useless unless
such a compilation
is of polynomial efficiency,
meaning that
its number of elementary
operations grows as poly($n$).

For concreteness,
we will use $n=3$
henceforth in this section,
but it will be obvious
how to draw
an analogous
circuit
for arbitrary $n$.

\begin{figure}[h]
    \begin{center}
    \epsfig{file=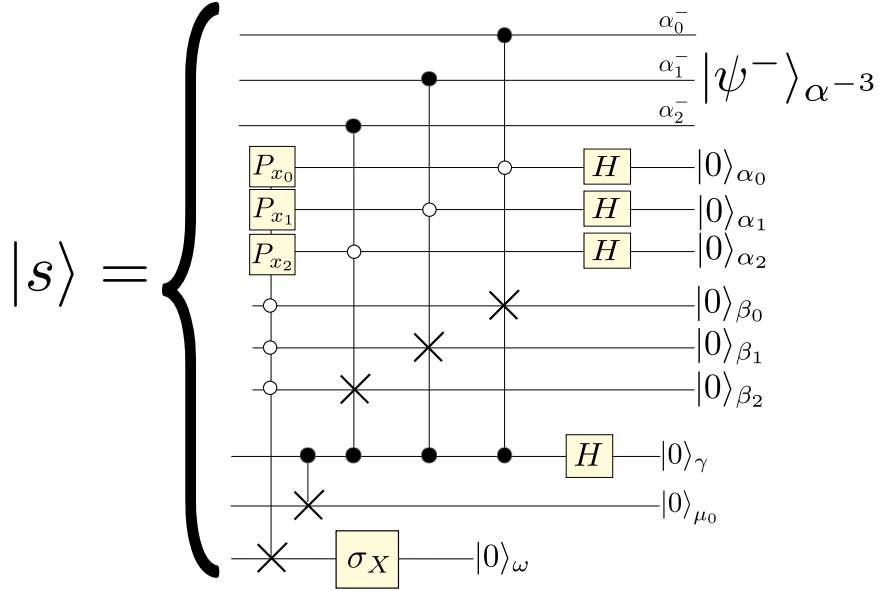, width=4.5in}
    \caption{Circuit for
    generating $\ket{s}$
    used in AFGA to calculate
    Mobius transform of $f^-(x^{-3})$.
    }
    \label{fig-qMob-ckt}
    \end{center}
\end{figure}

We want
all
horizontal lines
in Fig.\ref{fig-qMob-ckt}
to represent qubits.
Let
$\alpha^- = \alpha^{-3}$,
$\alpha = \alpha^3$, and
$\beta = \beta^3$.

Given $x^3\in Bool^3$, define

\beq
T(\alpha^-,\alpha,\beta)=
\prod_{j=0}^2
\left\{
\sigma_X(\beta_j)^{P_1(\alpha^-_j)
P_0(\alpha_j)}
H(\alpha_j)
\right\}
\;,
\eeq

\beq
\pi(\alpha)=
\prod_{j=0}^2
P_{x_j}(\alpha_j)
\;,
\eeq
and

\beq
\pi(\beta)=
\prod_{j=0}^2
P_{0}(\beta_j)
\;.
\eeq

Our method
for
calculating
the Mobius transform of $f^-(x^{-3})$
consists of applying the algorithm
AFGA\footnote{As discussed
 in Ref.\cite{qSym},
 we recommend the AFGA
 algorithm, but Grover's original
 algorithm (see Ref.\cite{Gro}) or any other
 Grover-like algorithm
 will also work
 here, as long as it
 drives a
 starting state $\ket{s}$
 to a target state $\ket{t}$.} of Ref.\cite{afga}
in the way that was described in
Ref.\cite{qSym},
using the techniques
of targeting two hypotheses
and blind targeting.
As in Ref.\cite{qSym},
when we apply AFGA in this section,
we will use a sufficient target $\ket{0}_\omega$.
All that remains for
us to do to
fully specify our
circuit for calculating
the Mobius transform of $f^-(x^{-3})$
is to give a circuit for
generating $\ket{s}$.

A circuit for generating
$\ket{s}$ is given by
Fig. \ref{fig-qMob-ckt}.
Fig.\ref{fig-qMob-ckt}
is equivalent to saying that

\beq
\ket{s}_{\mu,\nu,\omega}=
\sigma_X(\omega)^{
\pi(\beta)
\pi(\alpha)}
\frac{1}{\sqrt{2}}
\left[
\begin{array}{l}
T(\alpha^-,\alpha,\beta)
\begin{array}{l}
\ket{\psi^-}_{\alpha^{-}}
\\
\ket{0^3}_\alpha
\\
\ket{0^3}_\beta
\end{array}
\\
\ket{1}_\gamma
\\
\ket{1}_{\mu_0}
\\
\ket{1}_\omega
\end{array}
+
\begin{array}{l}
\ket{\psi^-}_{\alpha^{-}}
\\
H^{\otimes 3}\ket{0^3}_\alpha
\\
\ket{0^3}_\beta
\\
\ket{0}_\gamma
\\
\ket{0}_{\mu_0}
\\
\ket{1}_\omega
\end{array}
\right]
\;.
\eeq

\begin{claim}

\beq
\ket{s}_{\mu,\nu,\omega}=
\begin{array}{c}
z_1 \ket{\psi_1}_{\mu}
\\
\ket{1}_{\nu}
\\
\ket{0}_\omega
\end{array}
+
\begin{array}{c}
z_0 \ket{\psi_0}_{\mu}
\\
\ket{0}_{\nu}
\\
\ket{0}_\omega
\end{array}
+
\begin{array}{c}
\ket{\chi}_{\mu,\nu}
\\
\ket{1}_\omega
\end{array}
\;,
\eeq
for some unnormalized state
$\ket{\chi}_{\mu,\nu}$,
where

\beq
\begin{array}{|c|c|}
\hline
\ket{\psi_1}_{\mu}=
\frac{1}{\sqrt{f(x^3)}}
\sum_{x^{-3}}
\theta(x^3\geq x^{-3})
A^-(x^{-3})
\begin{array}{l}
\ket{x^{-3}}_{\alpha^-}
\\
\ket{x^3}_\alpha
\\
\ket{1}_{\mu_0}
\end{array}
&
\ket{\psi_0}_{\mu}=
\begin{array}{l}
\ket{\psi^-}_{\alpha^-}
\\
\ket{x^3}_\alpha
\\
\ket{0}_{\mu_0}
\end{array}
\\
\ket{1}_{\nu}=
\left[
\begin{array}{r}
\ket{0^3}_{\beta}
\\
\ket{1}_{\gamma}
\end{array}
\right]
&
\ket{0}_{\nu}=
\left[
\begin{array}{r}
\ket{0^3}_{\beta}
\\
\ket{0}_{\gamma}
\end{array}
\right]
\\
\hline
\end{array}
\;,
\eeq

\beq
z_1= \frac{1}{\sqrt{2^4}}
\sqrt{f(x^3)}
\;,
\eeq

\beq
z_0 =
\frac{1}{\sqrt{2^4}}
\;,
\eeq

\beq
\frac{|z_1|}{|z_0|} = \sqrt{\frac{P(1)}{P(0)}}
\;.
\eeq
\end{claim}
\proof

Recall that for any
quantum systems $\alpha$ and $\beta$,
any
unitary operator $U(\beta)$
and any
projection operator $\pi(\alpha)$,
one has

\beq
U(\beta)^{\pi(\alpha)}=
(1-\pi(\alpha)) + U(\beta)\pi(\alpha)
\;.
\label{eq-u-pi-id}
\eeq
Applying identity Eq.(\ref{eq-u-pi-id}) with $U=\sigma_X(\omega)$
yields:

\beqa
\ket{s} &=&
\sigma_X(\omega)^{\pi(\beta)\pi(\alpha)}\ket{s'}
\\
&=&
\sigma_X(\omega)\pi(\beta)\pi(\alpha)\ket{s'}
+
\begin{array}{l}
\ket{\chi}_{\mu,\nu}
\\
\ket{1}_\omega
\end{array}
\\
&=&
\frac{1}{\sqrt{2}}
\left[
\begin{array}{l}
\pi(\beta)
\pi(\alpha)T(\alpha^-,\alpha,\beta)
\begin{array}{l}
\ket{\psi^-}_{\alpha^{-}}
\\
\ket{0^3}_\alpha
\\
\ket{0^3}_\beta
\end{array}
\\
\ket{1}_\gamma
\\
\ket{1}_{\mu_0}
\\
\ket{0}_\omega
\end{array}
+
\begin{array}{l}
\ket{\psi^-}_{\alpha^-}
\\
\frac{1}{\sqrt{2^3}}\ket{x^3}_\alpha
\\
\ket{0^3}_\beta
\\
\ket{0}_\gamma
\\
\ket{0}_{\mu_0}
\\
\ket{0}_\omega
\end{array}
\right]
+
\begin{array}{l}
\ket{\chi}_{\mu,\nu}
\\
\ket{1}_\omega
\end{array}
\;.
\eeqa
Applying identity Eq.(\ref{eq-u-pi-id}) with $U=\sigma_X(\beta_j)$
yields:

\beqa
\lefteqn{
\pi(\beta)\pi(\alpha)
T(\alpha^-,\alpha,\beta)
\begin{array}{l}
\ket{\psi^-}_{\alpha^-}
\\
\ket{0^3}_\alpha
\\
\ket{0^3}_\beta
\end{array}
=}\nonumber
\\
&=&
\begin{array}{r}
\\
\\
\ket{0^3}_\beta
\end{array}
\sum_{x^{-3}}
\prod_{j=0}^2
\left\{
\begin{array}{r}
\\
P_{x_j}(\alpha_j)
\end{array}
\left[
\begin{array}{c}
\scriptstyle
1
-P_1(\alpha^-_j)P_0(\alpha_j)]
\\
\end{array}
\right]
\begin{array}{r}
\ket{x^-_j}_{\alpha^-_j}
\\
H(\alpha_j)\ket{0}_{\alpha_j}
\end{array}
\right\}
\begin{array}{l}
\av{x^{-3}|\psi^-}_{\alpha^-}
\\
\end{array}
\\
&=&
\begin{array}{r}
\\
\\
\ket{0^3}_\beta
\end{array}
\sum_{x^{-3}}
\begin{array}{l}
A^-(x^{-3})\ket{x^{-3}}_{\alpha^-}
\\
\ket{x^3}_\alpha
\end{array}
\prod_{j=0}^{2}
C(x^{-}_j,x_j)
\;,
\eeqa
where

\beqa
C(x^-_j,x_j)&=&
\begin{array}{l}
\bra{x^-_j}_{\alpha^-_j}
\\
\bra{x_j}_{\alpha_j}
\end{array}
[1-P_1(\alpha^-_j)P_0(\alpha_j)]
\begin{array}{r}
\ket{x^-_j}_{\alpha^-_j}
\\
H(\alpha_j)\ket{0}_{\alpha_j}
\end{array}
\\
&=&
\frac{1}{\sqrt{2}}
\begin{array}{l}
\bra{x^-_j}_{\alpha^-_j}
\\
\bra{x_j}_{\alpha_j}
\end{array}
[1-P_1(\alpha^-_j)P_0(\alpha_j)]
\begin{array}{l}
\ket{x^-_j}_{\alpha^-_j}
\\
\ket{x_j}_{\alpha_j}
\end{array}
\\
&=&
\frac{1}{\sqrt{2}}
\theta(x_j\geq x^-_j)
\;.
\eeqa
\qed

\section{Finding Minimum Value
Using Algorithm For Mobius Transforms}

Previous papers (see Refs.\cite{Gro,Bra1,Durr,Bra2})
have proposed algorithms for
finding the minimum value of a function
via Grover's algorithm.
In this section, we give an alternative method
of doing this
 that is based on the
 just described
method for calculating  Mobius transforms.

Suppose $x^n, y^n\in Bool^n$, and $E(x^n)>0$ is the
function we wish to minimize.
Define a secondary function $D^-()$ which
is sharply peaked (a sort of
Dirac delta function) at the minimum
of the function $E()$.
For example, define

\beq
D^-(x^n) = \frac{\exp\left\{\beta \sum_{y^n} [E(y^n)-E(x^n)]\right\}}
{\sum_{x^n} num}
\;
\eeq
for some large enough positive
$\beta$.
If $E(x^n)$ is minimum
 when $x^n=X^n$, then
 assume $D^-(x^n)$ is almost equal to
 the Kronecker delta function $\delta(x^n,X^n)$.
Let $D()$ denote the Mobius
transform of $D^-()$.
Let's speak in terms of the
decimal representation $x=dec(x^n)$ of
 the points $x^n\in Bool^n$.
 Call $X$ the minimum of $E(x)$.
Assume $n=5$ for concreteness.
The domain of the function $D^-$
is $\{0,1,\dots,31\}$.
Calculate $D(X_0)$
with $X_0=15$.
If $D(15)$ is much smaller than 1, then that means that the peak  $X$
 is in $\{16, 17,\ldots,31\}$
so set $X_1=23$, the midpoint
of $\{16, 17, \ldots, 31\}$.
Otherwise, if
$D(15)$ is close to 1,
then that means that the peak $X$
 is in $\{0,1,\ldots,15\}$
so set $X_1=7$, the midpoint
of $\{0, 1 ,\ldots, 15\}$.
Repeating this
procedure, one gets a finite sequence
$X_0, X_1,X_2,\ldots$
that converges to the peak $X$.
We are simply performing a binary search
for $X$.

Of course, for large $n$,
this technique
for finding minima
is only useful
if  $\ket{\psi_D^-}$
(where $\sqrt{D^-(x^n)} = \av{x^n|\psi_D^-}$)
can be compiled into a SEO
of poly($n$) length.

\section{Quantum Circuit For
Calculating Marginal Probability Distributions}

In this section,
we will give a quantum
circuit for calculating
the marginal probability
distribution $P(y^{n_0})$ of a given joint probability
distribution $P(y^n)$,
where $n>n_0>0$ and $y^n=(y^{n-n_0}, y^{n_0}) \in Bool^n$.

When using
a Classical Bayesian network (CB net)
with nodes $\ul{V}=\{\rvv_0, \rvv_1, \ldots,
\rvv_{m}\}$, one is
often interested in finding
$P(Y|X)$, where $\ul{Y}$
and $\ul{X}$ are two disjoint subsets of $\ul{V}$.
$P(Y|X)$  is the ratio of $P(Y,X)$ and $P(X)$, which are two
marginal probability distributions
of the probability distribution
$P(V)$ for the full CB net.
Furthermore,
if node $\rvv_j$ has $N_{\rvv_j}$ states,
those states can be identified
with
distinct bit strings
of length
approximately $\log_2(N_{\rvv_j})$.
So we see that the task of
calculating $P(Y|X)$
for a CB net
reduces to the task
that we are considering in
this section,
calculating the marginals
of a probability distribution
$P(y^n)$, where $y^n\in Bool^n$.

Suppose $n, n_0$ are integers such that $n> n_0>0$. For $x^{-n}\in Bool^n$
and a {\it normalized} $n$-qubit state $\ket{\psi^-}$,
define

\beq
\ket{\psi^-}_{\alpha^{-n}}=
\sum_{x^{-n}}
A^-(x^{-n})\ket{x^{-n}}_{\alpha^{-n}}
\;,
\eeq

\beq
P(x^{-n}) = |A^-(x^{-n})|^2
\;,
\eeq

\beq
P(x^{n_0}) =\sum_{x^{-n}}
\theta(x^{n_0}=x^{-n_0})
P(x^{-n})
\;.
\eeq

We will assume that
we know how to compile
$\ket{\psi^-}_{\alpha^{-n}}$
(i.e., that
we can construct it starting
from $\ket{0^n}_{\alpha^{-n}}$
using a sequence of
elementary operations.
Elementary operations are
operations that act on a few (usually 1,2 or 3)
qubits at a time,
such as qubit rotations
and CNOTS.)
Multiplexor techniques for doing
such compilations
are discussed in Ref.\cite{tuc-multiplexor}.
If $n$
is very large,
our algorithm will
be useless unless
such a compilation
is of polynomial efficiency,
meaning that
its number of elementary
operations grows as poly($n$).

For concreteness,
we will use $n_0=3$
and $n$ arbitrary (but greater than $n_0$)
henceforth in this section,
but it will be obvious
how to draw
an analogous
circuit
for arbitrary $n_0$.

\begin{figure}[h]
    \begin{center}
    \epsfig{file=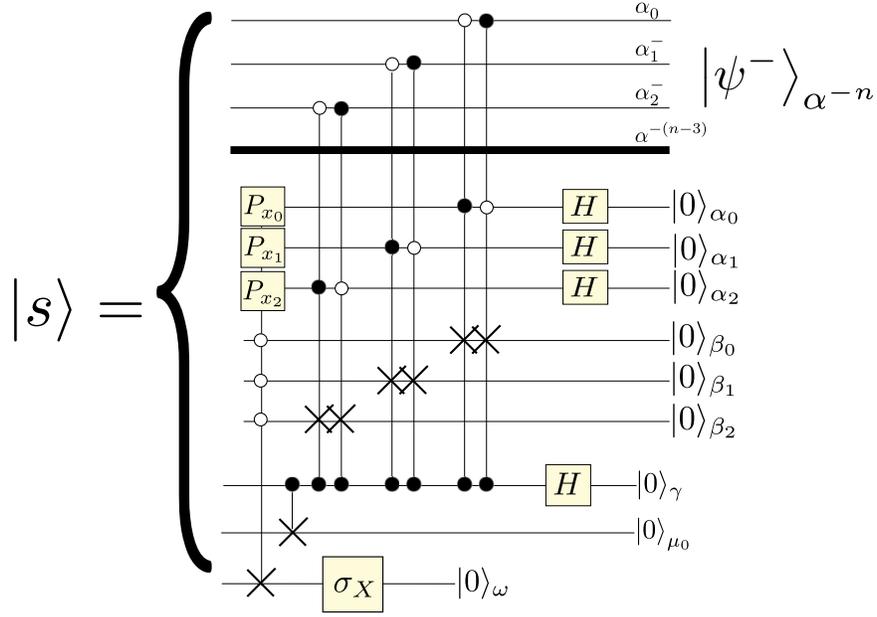, width=4.5in}
    \caption{Circuit for
    generating $\ket{s}$
    used in AFGA to calculate
    the marginal
    $P(x^{-3})$
of $P(x^{-n})$
evaluated at
$x^{-3}=x^3$.
    }
    \label{fig-qMargi-ckt}
    \end{center}
\end{figure}

We want
all
horizontal lines
in Fig.\ref{fig-qMargi-ckt}
to represent qubits,
except for the thick line labelled
$\alpha^{-(n-3)}$ which represents $n-3$ qubits.
Let $\alpha^- = \alpha^{-n}$,
$\alpha = \alpha^3$, and
$\beta = \beta^3$.
Note that in the qMobius case,
 the number of $\alpha^-$, $\alpha$, $\beta$ qubits were all the same,
 whereas
in this case, there are $n$ $\alpha^-$ qubits but only $3$ $\alpha$ and $\beta$ ones.

Given $x^3\in Bool^3$, define

\beq
T(\alpha^-,\alpha,\beta)=
\prod_{j=0}^2
\left\{
\sigma_X(\beta_j)^{P_1(\alpha^-_j)
P_0(\alpha_j)+ P_0(\alpha^-_j)
P_1(\alpha_j) }
H(\alpha_j)
\right\}
\;,
\eeq

\beq
\pi(\alpha)=
\prod_{j=0}^2
P_{x_j}(\alpha_j)
\;,
\eeq
and

\beq
\pi(\beta)=
\prod_{j=0}^2
P_{0}(\beta_j)
\;.
\eeq

Our method
for
calculating
the marginal
$P(x^{-3})$
of $P(x^{-n})$
evaluated at
$x^{-3}=x^3$
consists of applying the algorithm
AFGA\footnote{As discussed
 in Ref.\cite{qSym},
 we recommend the AFGA
 algorithm, but Grover's original
 algorithm (see Ref.\cite{Gro}) or any other
 Grover-like algorithm
 will also work
 here, as long as it
 drives a
 starting state $\ket{s}$
 to a target state $\ket{t}$.} of Ref.\cite{afga}
in the way that was described in
Ref.\cite{qSym},
using the techniques
of targeting two hypotheses
and blind targeting.
As in Ref.\cite{qSym},
when we apply AFGA in this section,
we will use a sufficient target $\ket{0}_\omega$.
All that remains for
us to do to
fully specify our
circuit for calculating
$P(x^3)$
is to give a circuit for
generating $\ket{s}$.

A circuit for generating
$\ket{s}$ is given by
Fig. \ref{fig-qMargi-ckt}.
Fig.\ref{fig-qMargi-ckt}
is equivalent to saying that

\beq
\ket{s}_{\mu,\nu,\omega}=
\sigma_X(\omega)^{
\pi(\beta)
\pi(\alpha)}
\frac{1}{\sqrt{2}}
\left[
\begin{array}{l}
T(\alpha^-,\alpha,\beta)
\begin{array}{l}
\ket{\psi^-}_{\alpha^{-}}
\\
\ket{0^3}_\alpha
\\
\ket{0^3}_\beta
\end{array}
\\
\ket{1}_\gamma
\\
\ket{1}_{\mu_0}
\\
\ket{1}_\omega
\end{array}
+
\begin{array}{l}
\ket{\psi^-}_{\alpha^{-}}
\\
H^{\otimes 3}\ket{0^3}_\alpha
\\
\ket{0^3}_\beta
\\
\ket{0}_\gamma
\\
\ket{0}_{\mu_0}
\\
\ket{1}_\omega
\end{array}
\right]
\;.
\eeq

\begin{claim}

\beq
\ket{s}_{\mu,\nu,\omega}=
\begin{array}{c}
z_1 \ket{\psi_1}_{\mu}
\\
\ket{1}_{\nu}
\\
\ket{0}_\omega
\end{array}
+
\begin{array}{c}
z_0 \ket{\psi_0}_{\mu}
\\
\ket{0}_{\nu}
\\
\ket{0}_\omega
\end{array}
+
\begin{array}{c}
\ket{\chi}_{\mu,\nu}
\\
\ket{1}_\omega
\end{array}
\;,
\eeq
for some unnormalized state
$\ket{\chi}_{\mu,\nu}$,
where

\beq
\begin{array}{|c|c|}
\hline
\ket{\psi_1}_{\mu}=
\frac{1}{\sqrt{P(x^3)}}
\sum_{x^{-n}}
\theta(x^3= x^{-3})
A^-(x^{-n})
\begin{array}{l}
\ket{x^{-n}}_{\alpha^-}
\\
\ket{x^3}_\alpha
\\
\ket{1}_{\mu_0}
\end{array}
&
\ket{\psi_0}_{\mu}=
\begin{array}{l}
\ket{\psi^-}_{\alpha^-}
\\
\ket{x^3}_\alpha
\\
\ket{0}_{\mu_0}
\end{array}
\\
\ket{1}_{\nu}=
\left[
\begin{array}{r}
\ket{0^3}_{\beta}
\\
\ket{1}_{\gamma}
\end{array}
\right]
&
\ket{0}_{\nu}=
\left[
\begin{array}{r}
\ket{0^3}_{\beta}
\\
\ket{0}_{\gamma}
\end{array}
\right]
\\
\hline
\end{array}
\;,
\eeq

\beq
z_1= \frac{1}{\sqrt{2^4}}
\sqrt{P(x^3)}
\;,
\eeq

\beq
z_0 =
\frac{1}{\sqrt{2^4}}
\;,
\eeq

\beq
\frac{|z_1|}{|z_0|} = \sqrt{\frac{P(1)}{P(0)}}
\;.
\eeq
\end{claim}
\proof

Recall that for any
quantum systems $\alpha$ and $\beta$,
any
unitary operator $U(\beta)$
and any
projection operator $\pi(\alpha)$,
one has

\beq
U(\beta)^{\pi(\alpha)}=
(1-\pi(\alpha)) + U(\beta)\pi(\alpha)
\;.
\label{eq-u-pi-id-margi}
\eeq
Applying identity Eq.(\ref{eq-u-pi-id-margi}) with $U=\sigma_X(\omega)$
yields:

\beqa
\ket{s} &=&
\sigma_X(\omega)^{\pi(\beta)\pi(\alpha)}\ket{s'}
\\
&=&
\sigma_X(\omega)\pi(\beta)\pi(\alpha)\ket{s'}
+
\begin{array}{l}
\ket{\chi}_{\mu,\nu}
\\
\ket{1}_\omega
\end{array}
\\
&=&
\frac{1}{\sqrt{2}}
\left[
\begin{array}{l}
\pi(\beta)
\pi(\alpha)T(\alpha^-,\alpha,\beta)
\begin{array}{l}
\ket{\psi^-}_{\alpha^{-}}
\\
\ket{0^3}_\alpha
\\
\ket{0^3}_\beta
\end{array}
\\
\ket{1}_\gamma
\\
\ket{1}_{\mu_0}
\\
\ket{0}_\omega
\end{array}
+
\begin{array}{l}
\ket{\psi^-}_{\alpha^{-}}
\\
\frac{1}{\sqrt{2^3}}\ket{x^3}_\alpha
\\
\ket{0^3}_\beta
\\
\ket{0}_\gamma
\\
\ket{0}_{\mu_0}
\\
\ket{0}_\omega
\end{array}
\right]
+
\begin{array}{l}
\ket{\chi}_{\mu,\nu}
\\
\ket{1}_\omega
\end{array}
\;.
\eeqa
Applying identity Eq.(\ref{eq-u-pi-id-margi}) with $U=\sigma_X(\beta_j)$
yields:

\beqa
\lefteqn{
\pi(\beta)\pi(\alpha)
T(\alpha^-,\alpha,\beta)
\begin{array}{l}
\ket{\psi^-}_{\alpha^-}
\\
\ket{0^3}_\alpha
\\
\ket{0^3}_\beta
\end{array}
=}\nonumber
\\
&=&
\sum_{x^{-n}}
\begin{array}{l}
\\
\scriptstyle
\ket{x^{-(n-3)}}_{\alpha^{-(n-3)}}
\\
\scriptstyle
\ket{0^3}_\beta
\end{array}
\prod_{j=0}^2
\left\{
\begin{array}{r}
\\
\scriptstyle
 P_{x_j}(\alpha_j)
\end{array}
\left[
\begin{array}{r}
\scriptstyle
1
-P_0(\alpha^-_j)P_1(\alpha_j)
\\
\scriptstyle
-P_1(\alpha^-_j)P_0(\alpha_j)
\end{array}
\right]
\begin{array}{r}
\scriptstyle
\ket{x^-_j}_{\alpha^-_j}
\\
\scriptstyle
H(\alpha_j)\ket{0}_{\alpha_j}
\end{array}
\right\}
\scriptstyle
\av{x^{-n}|\psi^-}_{\alpha^-}
\\
&=&
\begin{array}{r}
\\
\\
\ket{0^3}_\beta
\end{array}
\sum_{x^{-n}}
\begin{array}{l}
A^-(x^{-n})\ket{x^{-n}}_{\alpha^-}
\\
\ket{x^3}_\alpha
\end{array}
\prod_{j=0}^{2}
C(x^{-}_j,x_j)
\;,
\eeqa
where

\beqa
C(x^-_j,x_j)&=&
\begin{array}{l}
\bra{x^-_j}_{\alpha^-_j}
\\
\bra{x_j}_{\alpha_j}
\end{array} [P_1(\alpha^-_j)P_1(\alpha_j)
+P_0(\alpha^-_j)P_0(\alpha_j)]
\begin{array}{r}
\ket{x^-_j}_{\alpha^-_j}
\\
H(\alpha_j)\ket{0}_{\alpha_j}
\end{array}
\\
&=&
\frac{1}{\sqrt{2}}
\begin{array}{l}
\bra{x^-_j}_{\alpha^-_j}
\\
\bra{x_j}_{\alpha_j}
\end{array}
[P_1(\alpha^-_j)P_1(\alpha_j)
+P_0(\alpha^-_j)P_0(\alpha_j)]
\begin{array}{l}
\ket{x^-_j}_{\alpha^-_j}
\\
\ket{x_j}_{\alpha_j}
\end{array}
\\
&=&
\frac{1}{\sqrt{2}}
\theta(x_j= x^-_j)
\;.
\eeqa
\qed


\begin{thebibliography}{99}
\bibitem{KS}
R. Kennes, P. Smets, ``Computational aspects of the Mobius transform",  arXiv:1304.1122

\bibitem{Ko-So}
M. Koivisto, and K. Sood, ``Exact Bayesian structure discovery in Bayesian networks",  The Journal of Machine Learning Research 5 (2004): 549-573.

\bibitem{Gro}Lov K. Grover,
``Quantum computers can search
rapidly by using almost any transformation",
arXiv:quant-ph/9712011

\bibitem{afga}
R.R. Tucci, ``An Adaptive, Fixed-Point Version of Grover's Algorithm", arXiv:1001.5200

\bibitem{qSym}
R.R. Tucci,
``Quantum Circuit for Calculating
Symmetrized Functions
Via Grover-like Algorithm", arXiv:1403.6707

\bibitem{tuc-multiplexor}
R.R. Tucci, ``Code Generator for Quantum Simulated Annealing", arXiv:0908.1633

\bibitem{Bra1}
G. Brassard, P. Hoyer, M. Mosca, and A. Tapp, ``Quantum amplitude amplification and estimation",  arXiv:quant-ph/0005055



\bibitem{Durr}
C. D\"{u}rr, P. Hoyer, ``A quantum algorithm for finding the minimum",  arXiv:quant-ph/9607014.

\bibitem{Bra2}
G. Brassard, F. Dupuis, S. Gambs, and A. Tapp, ``An optimal quantum algorithm to approximate the
mean and its application for approximating the median of a set of points over an arbitrary distance",  arXiv:1106.4267

\end{thebibliography}
\end{document}